\documentclass[twocolumn,showpacs,aps,prl,superscriptaddress]{revtex4}

\usepackage{amsmath}
\usepackage{epsfig}

\input babarsym

\newcommand{\BABARPubYear}    {10}
\newcommand{\BABARPubNumber} {010}

\newcommand{\SLACPubNumber}{SLAC-PUB-14245}

\def\figurebox#1#2#3{%
    \def\arg{#3}%
    \ifx\arg\empty
    {\hfill\vbox{\hsize#2\hrule\hbox to #2{\vrule\hfill\vbox to #1{\hsize#2\vfill}\vrule}\hrule}\hfill}%
    \else
    {\hfill\epsfbox{#3}\hfill}%
    \fi}

\begin{document}

\preprint{\babar-PUB-\BABARPubYear/\BABARPubNumber}
\preprint{SLAC-PUB-\SLACPubNumber} 

\begin{flushleft}
\babar-PUB-\BABARPubYear/\BABARPubNumber\\
SLAC-PUB-\SLACPubNumber\\
\end{flushleft}

\title{
{\large \bf
  Observation of the Decay \boldBtoDsKlnuX }
}

%
\author{P.~del~Amo~Sanchez}
\author{J.~P.~Lees}
\author{V.~Poireau}
\author{E.~Prencipe}
\author{V.~Tisserand}
\affiliation{Laboratoire d'Annecy-le-Vieux de Physique des Particules (LAPP), Universit\'e de Savoie, CNRS/IN2P3,  F-74941 Annecy-Le-Vieux, France}
\author{J.~Garra~Tico}
\author{E.~Grauges}
\affiliation{Universitat de Barcelona, Facultat de Fisica, Departament ECM, E-08028 Barcelona, Spain }
\author{M.~Martinelli$^{ab}$}
\author{A.~Palano$^{ab}$ }
\author{M.~Pappagallo$^{ab}$ }
\affiliation{INFN Sezione di Bari$^{a}$; Dipartimento di Fisica, Universit\`a di Bari$^{b}$, I-70126 Bari, Italy }
% ...
\collaboration{The \babar\ Collaboration}
\noaffiliation

\date{\today}


\begin{abstract}
We report the observation of the decay \BtoDsKlnuX\ based on $342 \invfb$ of data
collected at the \FourS\ resonance with the \babar\ detector at the \pep2\ \epem\ storage rings at SLAC.
A simultaneous fit to three \Ds\ decay chains is performed to extract the signal yield from 
measurements of the squared missing mass in the \B\ meson decay. 
We observe the decay $\BtoDsKlnuX$ with a significance greater than five standard deviations 
(including systematic uncertainties) and measure its branching fraction to be
$\BR(\BtoDsKlnuX) = [6.13^{+1.04}_{-1.03}(\mathrm{stat.})\pm0.43(\mathrm{syst.}) \pm 0.51(\BR(\DsO))]\times10^{-4}$, where  the last error reflects the limited knowledge of the \DsO\ branching fractions.

\end{abstract}
\pacs{13.20.He, 12.15.Ji, 12.38.Qk}
\maketitle

The study of charmed inclusive semileptonic $B$ meson decays enables the measurement of the 
CKM matrix element $|V_{cb}|$. This measurement relies on a precise knowledge of all semileptonic \B\ meson 
decays. Decays of orbitally excited $D$ mesons, from the process $B \to D^{**}\ell \nu$, constitute a significant fraction of these decays 
\cite{PDG}, and may help explain the discrepancy between the inclusive $B \to X_c\ell \nu$ 
rate, where $X_c$ is a charmed hadronic final state, and the sum of the measured exclusive decay rates \cite{PDG, InclExclComparison}. 
So far, analyses of these decays have focused on the reconstruction of 
$B \to D^{(*)}\pi \ell \nu$ states \cite{DLP, Narrow1, Narrow2}. In such analyses, experimental data are interpreted as a sum of the four $D^{**}$ resonances. The results show the dominance of $B$ decays to broad resonances, while QCD sum rules imply the opposite \cite{Broad}.  Conversely, a small contribution from broad $D^{**}$ states implies the presence
of a non-resonant $B \to D^{(*)}\pi \ell \nu$ component, which has not yet been observed. Measurement of the branching fraction for the as-yet-unobserved $\BtoDsKlnuX$ decay \cite{conjugate} would provide additional information relevant to this issue, by exploring the hadronic mass distribution above $2.46\gevcc$ where resonant and non-resonant components are present. In addition, the measurement of $\BtoDsKlnuX$ will provide a better estimate of background in future studies of semileptonic $\BstoDsXlnu$ decays. 

Using the shape of the hadronic mass spectrum in $B$ semileptonic decays, a rough estimate on the branching fraction $\BR(\BtoDsKlnuX)$ is of the order of $10^{-3}$ \cite{Bigi, LEP}, which is consistent with the limit set by the ARGUS Collaboration, $\BR(\BtoDsKlnuX) < 5\times 10^{-3}$ at $90\%$ confidence level \cite{Argus}. A comparison between this expectation and the actual measurement can confirm or refute the expected rapid decrease of the hadronic mass distribution at high values.

In this paper, we present the observation of $\BtoDsKlnuX$ decays, where $\ell = e,\, \mu$. This analysis does not differentiate between final states with \Ds\ and \Dss, where $\Dss$ decays via emission of neutral decay products that are not reconstructed. 
The results are based on a data sample of $N_{\BB} = (376.9\pm 4.1)\times 10^{6}$ \BB\ pairs recorded at the \FourS\ resonance with the \babar\ detector \cite{BabarDet} at the \pep2\ asymmetric energy \epem\ storage rings at the SLAC National Accelerator Laboratory. This  corresponds to an integrated luminosity of  $342 \invfb$. In addition, $37 \invfb$ of data 
collected about $40 \mev$ below the resonance are used for background studies. A \geant-based Monte Carlo
(MC) simulation \cite{Geant} of \BB\ and continuum events ($\epem \to \qqbar$ with $\q = \u,\d,\s,\c$) is used to 
study the detector response and acceptance, validate the analysis technique, and evaluate signal efficiencies.
The sample of simulated \BB\ events is equivalent to approximately 3 times the data sample. The signal MC events are generated by adapting the decay model of Goity and Roberts \cite{GoiRob} to describe $\Ds\Km$ final states. 
Two alternative signal MC samples are used to estimate systematic uncertainties: a sample based on the ISGW2 model \cite{ISGW2}, in which \Bm\ mesons decay to $\Dstarz_0 \ellm \nub$ with $\Dstarz_0 \to \Ds \Km$, and a sample based on a simple phase space model. The signal MC samples are equivalent to approximately 10 times the expected signal yield. 


We reconstruct $\Ds$ candidates in three decay chains: 
$\Ds \to \phi \pip$ with $\phi \to \Kp \Km$, $\Ds \to \Kstarzb \Kp$ with $\Kstarzb \to \Km \pip$, and $\Ds \to \KS \Kp$
with $\KS \to \pip \pim$. The $\phi$, $\Kstarzb$, and $\KS$ candidates are formed by combining oppositely charged tracks. To suppress combinatorial background from the \Ds\ reconstruction in the first two decay chains, we employ a feed-forward neural network (multilayer perceptron, MLP \cite{TMVA}) with three input variables and four hidden layers. The input variables are the
absolute value of the difference between the reconstructed and the nominal mass values of the  $\phi/\Kstarzb$ candidate \cite{PDG}, the absolute value of the cosine of the helicity angle of the $\phi/\Kstarzb$, and the 
$\chi^2$ probability of the fit to the $\Ds$ candidate. The helicity angle is defined as the angle
between the $\Ds$ candidate and one kaon originating from the $\phi/\Kstarzb$ in the $\phi/\Kstarzb$ rest frame. To
suppress combinatorial background in the $\Ds \to \KS \Kp$ decay chain, we require the invariant mass of the charged pions forming the $\KS$ candidate
to satisfy $0.490 \gevcc < m(\pi\pi) < 0.506 \gevcc$, the flight length of the $\KS$ to be larger than $1\mm$, the 
cosine of the laboratory angle between the $\KS$ momentum and the line connecting the $\KS$ decay vertex and the primary 
vertex of the event to be positive, and the probability of the $\Ds$ candidate's vertex fit to be 
larger than $0.001$. The selection criteria are optimized to maximize the statistical significance of the signal. No requirement on the mass of the 
\Ds\ candidates is applied, since this distribution is used to extract the signal yield.

A lepton and a kaon, both with negative charge, are combined with the \Ds\ candidate to form a \Bm\ candidate. Leptons are required to have momentum $|\vec{p}_\ell|$ larger than $0.8\gevc$ \cite{CMframe} to reject those not directly originating from $B$ mesons. The probability of the vertex fit of the \B\ candidate is required to be larger than $0.01$. 

Three event-shape variables that are sensitive to the topological differences between jet-like continuum events and more 
spherical \BB\ events are used as input to a neural network to suppress background from continuum events. These variables are the normalized second Fox-Wolfram moment $R_2$ \cite{FWMoments}, 
the monomial $L_2$ \cite{Legendre}, and the cosine of the angle between the flight direction of the reconstructed \B\  candidate and the rest of the event. 
A neural network whose input variables are the \B\ candidate mass, the \B\ candidate sphericity, and the thrust value of the rest of the event, is used to reduce the background from other \B\ decays, providing a slight, but not negligible improvement in the sensitivity of the measurement.

\begin{figure}[!hbt]
  \centering
  \includegraphics[width=0.238\textwidth]{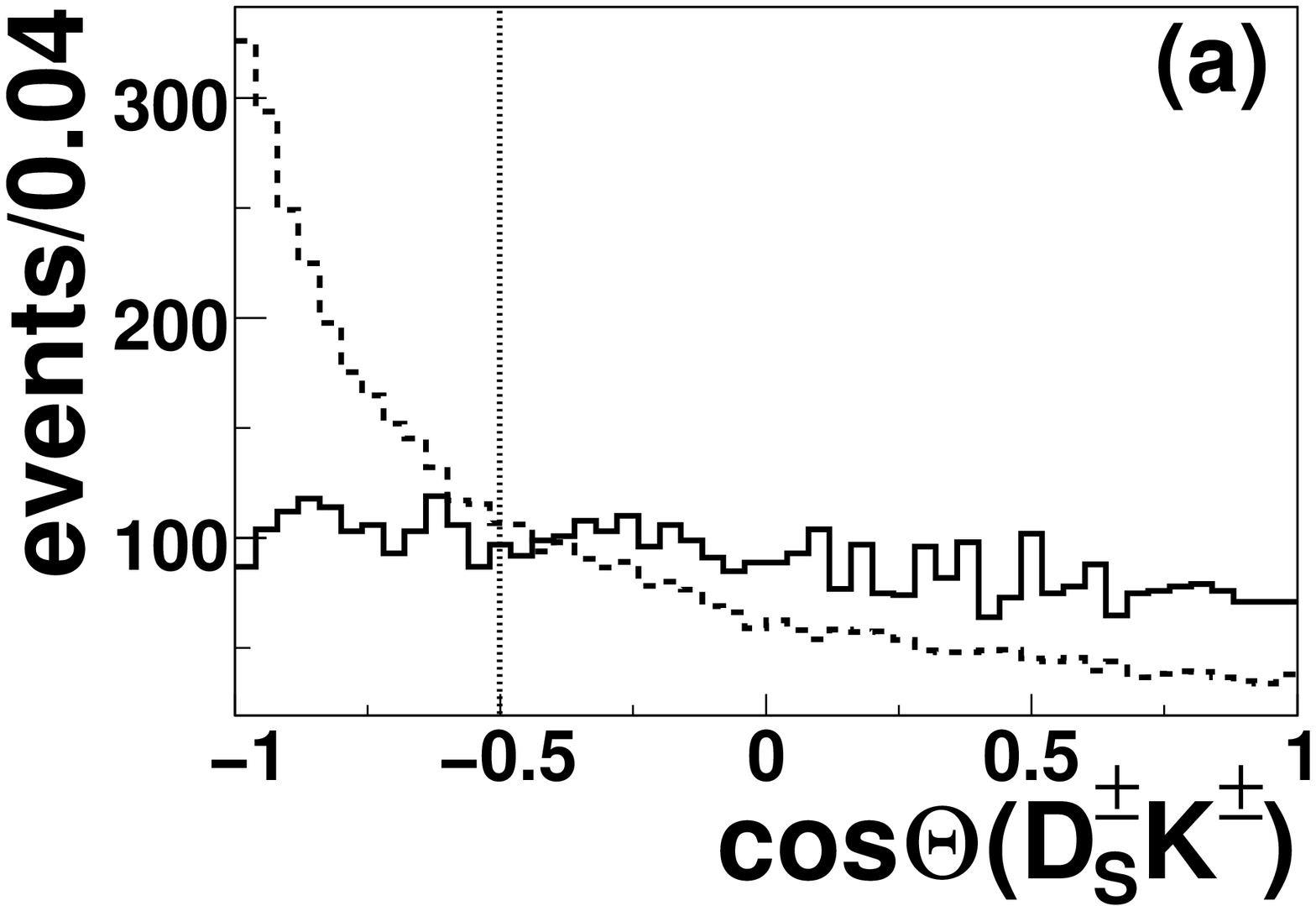}
  \begin{minipage}[b]{0.238\textwidth}
    \includegraphics[width=\textwidth]{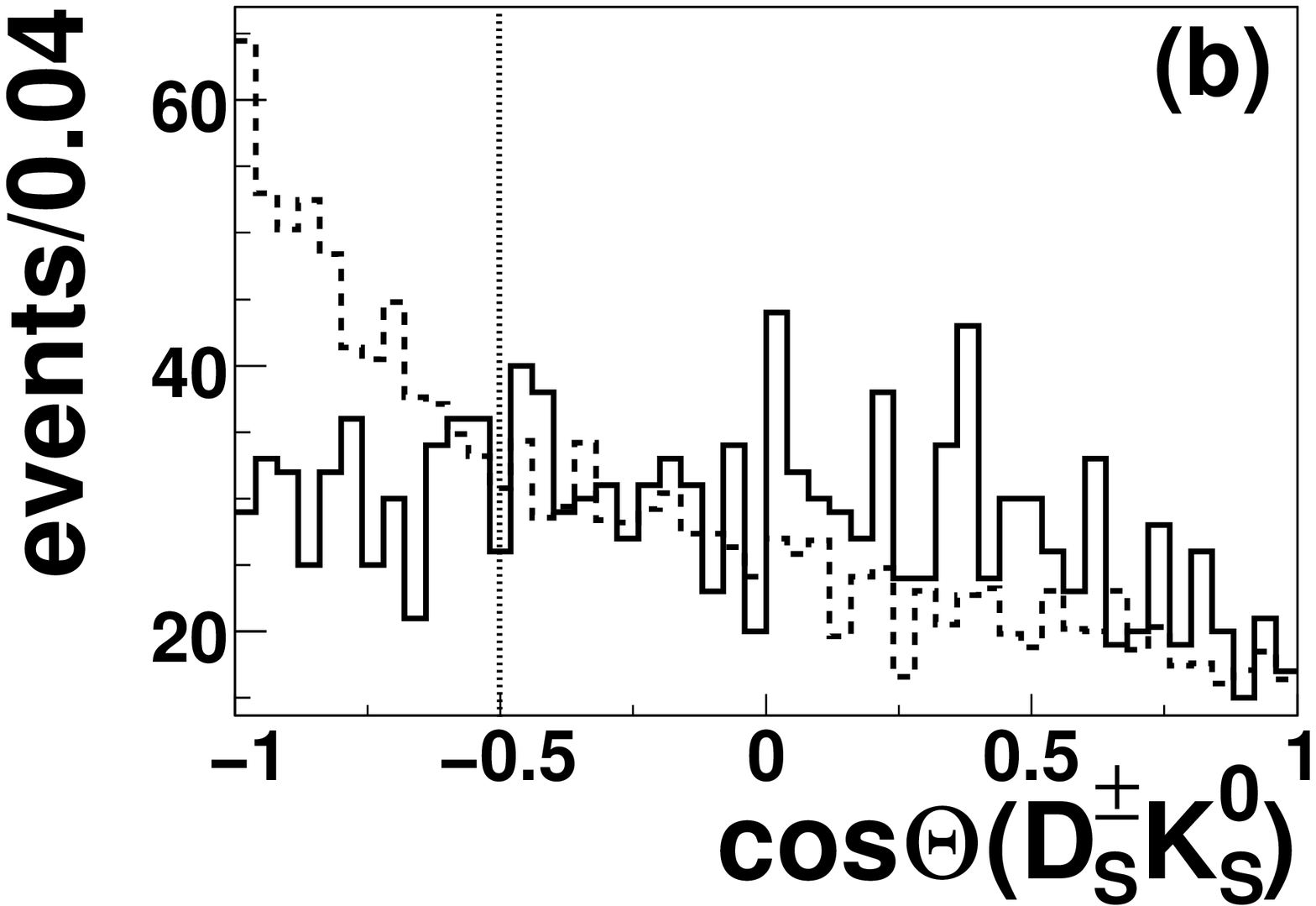}
  \end{minipage}\\[-8pt]
  \caption[Angular distributions of $cos\Theta_{\Ds K}$]
  {\sl Angular distribution of the cosine of the angle between the flight direction of the \Dspm\ meson and additional charged and 
    neutral kaons: (a) $\cos\Theta(\Dspm K^\pm)$ and (b) $\cos\Theta(\Dspm \KS)$. Solid lines represent signal MC events; dashed lines are \Ds\ background. The vertical lines indicate the selection applied.}
  \label{fig:DsKAngles}
\end{figure}

After applying these selection criteria, the remaining background events are divided into two classes, depending on whether or not they contain a correctly reconstructed \Ds\ meson. The first class is the more important of the two. We refer  to it as \Ds\ background events in the following. Most of these events contain a \DsO\ originating from decays such as $\B \to \DsO D$, where the kaon and lepton tracks 
used to form a \B\ candidate are taken from the other \B\ meson in the event. The angular correlation between the flight directions of the \DsO\ and the $D$ is used to suppress the \DsO\ background candidates. The direction of the $D$ meson is estimated from the direction of a previously unused charged or neutral kaon candidate that is assumed to be from $D \to \KpmO X$ decays.
Requiring the cosine of the angle between the flight direction of the \DsO\ candidate and the additional kaon to be 
larger than $-0.5$, about $30 \%$ of the \DsO-background events are rejected, as shown in 
Fig.~\ref{fig:DsKAngles}.
About $8\%$ of the remaining events have multiple candidates, predominantly two. In such cases, we choose the candidate with the largest $B$ vertex fit probability.

The remaining events are divided into signal regions and sidebands based on the mass of the
\Ds\ candidate. The sidebands are defined by $1.9\gevcc < m(\Ds) < 1.94\gevcc$ and 
$2.0\gevcc < m(\Ds) < 2.04\gevcc$. Fits to the \Ds\ mass distributions are performed separately for each decay channel to define the signal regions, and to measure the number of 
reconstructed \Ds\ mesons, which are used later for extracting the signal yield. The signal regions are defined as $\pm2.5\sigma$ wide bands, centered on the "fitted means" for each decay channel. Signal events are identified by the missing mass of the visible decay products $Y=\Ds \Km \ellm$ with respect to the
nominal $B$ meson mass:
\begin{equation}
    \MM = (E_{B} - E_{Y})^2 - |\vec{p}_{Y}|^{2} = m_{\nu}^{2},
\end{equation}
where $E_B$ is the beam energy, corresponding to the energy of the $B$ meson, while $E_Y$ and $\vec{p}_{Y}$ represent the energy and momentum of the $Y$ composite, respectively. Due to its smallness and unknown direction, the momentum of the $B$ meson is neglected. This leads to a distribution for \MM\ with a Gaussian shape for correctly reconstructed signal events. Other $B$ semileptonic decays, where one particle is not reconstructed or is erroneously included, lead to higher or lower values of \MM.

To extract the signal yield, we perform an unbinned extended maximum-likelihood fit, applied simultaneously to the \MM\ distributions of the signal region and the sidebands of the three \Ds\ decay chains. While the sidebands are populated only by combinatorial background events, the signal region also contains \Ds\ background and signal events. Because their lepton acceptances differ, the electron and muon channels are fitted separately. The combinatorial background is modeled using a sum of two Gaussian distributions whose parameters are the same for the three \Ds\ decay chains. This parameterization is favored by MC simulation. This fit technique is equivalent to a sideband subtraction. The 
contributions of \Ds\ background events are modeled using a Fermi function:
\begin{equation}
    f(\MM) = \frac{1}{e^{[(\MM-\MO)/\EC]} + 1},
\end{equation}
where \MO\ represents the \MM\ drop-off value and \EC\ the smearing of the Fermi-edge. The values
for \MO\ and \EC, $\MO=(0.303\pm0.034)\gevccc$ and $\EC =  (0.333\pm0.018) \gevccc$ for the electron channel, and $\MO=(0.247\pm0.041)\gevccc$ and $\EC =  (0.346\pm0.022) \gevccc$ for the muon channel, are fixed to the values derived from fits to MC distributions and are the same for all \Ds\ decay chains. Signal events are 
modeled by a Gaussian distribution, with the same mean and width for all reconstruction channels. The width is fixed to the value determined from the simulation. The mean of the distribution is determined in the fit, allowing for contributions from events with a \Dss\ in the final state.

The total number of events with a \Ds\ and the number of combinatorial background events in the signal region have been determined from fits to the $m(\Ds)$ distributions. The number of signal $\BtoDsKlnuX$ and \Ds\ background events are extracted from the fits to the \MM\ distributions, separately for the electron and muon samples. For these fits the three \Ds\ decay channels are combined, taking into account their detection branching fractions $\epsilon_{BR} = \BR(\Ds\to D1\,d2)\times \BR(D1\to d3\,d4)$ and individual reconstruction efficiencies $\epsilon_{\mathrm{reco}}$. For illustration, these efficiencies and the branching ratios are listed in table \ref{tab:fitresults}, together with the total fitted number of signal events and the estimated contributions from each of the three channels.

The fit is performed in the range $|\MM|<1.5 \gevccc$ has 10 free parameters: the mean value of \MM, the total number of fitted signal events $N^{\mathrm{signal}}$, five parameters that describe the shape of the combinatorial background, and three sideband normalization parameters. The number of signal and \Ds\ background events are free in the fit, only the sum of both values is constrained to the result of the fits to the $m(\Ds)$ distributions. 

The likelihood function is 
\begin{equation} 
  \mathcal{L}=\frac{e^{-N^{\mathrm{signal}}}}{n!}(N^{\mathrm{signal}})^{n}\prod_{j}\prod_i^{N_{j}} {\mathcal{P}}(M_{m,i}^{2},\, \alpha_{j}),
\end{equation}
with $N_{j}$ the number of events and ${\mathcal{P}}({M_{m,i}^{2}}\, \alpha_{j})$ the probability density function (PDF) for a given fit slice $j$ (signal region or sideband of each \Ds\ decay chain) with the fit parameters $\alpha$, and $n = \sum_{j} N_{j}$ the total number of events. 

\begin{table*}[hbt]
  \begin{center}
    \caption{Signal yields, selection efficiencies $\epsilon_{\mathrm{reco}}$, and branching fractions $\epsilon_{BR} = \BR(\Ds\to D1\,d2)\times \BR(D1\to d3\,d4)$ for the individual and combined decay chains. The signal yields of each decay chain are computed using $N^{\mathrm{signal}}$ and the efficiencies and are given for illustration only. The errors on the signal yields are the fit errors, the uncertainties of $\epsilon_{\mathrm{reco}}$ are the systematic uncertainties and the uncertainties of $\epsilon_{BR}$ represent the limited knowledge of the branching fractions of the \Ds.}
    \vspace{2pt}
    \label{tab:fitresults}
    \begin{tabular}{l c c c c c}
      \hline \hline
      \Ds\ decay chain & \bf $N^{\mathrm{signal}}_{\mathrm{electron}}$ & $\epsilon_{\mathrm{reco,\, electron}} \; [\%]$ & $N^{\mathrm{signal}}_{\mathrm{muon}}$ & $\epsilon_{\mathrm{reco, \, muon}}\; [\%]$ & $\epsilon_{BR}\; [\%]$ \\
      \hline
      all & $259.4^{+67.6}_{-67.2}$ & & $209.7^{+53.0}_{-52.2}$ & & \\
      \hline
      $\Ds \to \phi \pip,\, \phi \to \Kp \Km$           & $115.7^{+30.2}_{-30.0}$ & $(2.76\pm0.08)$ & $92.1^{+23.3}_{-22.9}$ & $(1.62\pm0.06)$ & $(2.18\pm0.33)$\\
      $\Ds \to \Kstarzb \Kp,\, \Kstarzb \to \Km \pip$ & $85.2^{+22.2}_{-22.1}$ & $(1.79\pm0.06)$ & $70.2^{+17.8}_{-17.5}$ & $(1.09\pm0.05)$ & $(2.60\pm0.40)$\\
      $\Ds \to \KS \Kp,\, \KS \to \pip \pim$            & $58.5^{+15.3}_{-15.2}$ & $(2.98\pm0.08)$ & $47.4^{+12.0}_{-11.8}$ & $(1.78\pm0.06)$ & $(1.02\pm0.09)$\\
      \hline \hline
    \end{tabular}
  \end{center}
\end{table*}

Using MC experiments from a generator, which includes parameterizations of detector performance for signal reconstruction and background expectations, it has been verified that the fit is able to extract signal branching fractions for $\BR(\BtoDsKlnuX) > 3 \times 10^{-4}$. Values of fit biases are also determined with this procedure and are taken into account in the analysis.

\begin{figure}[!hbt]
  \centering
  \includegraphics[width=0.238\textwidth]{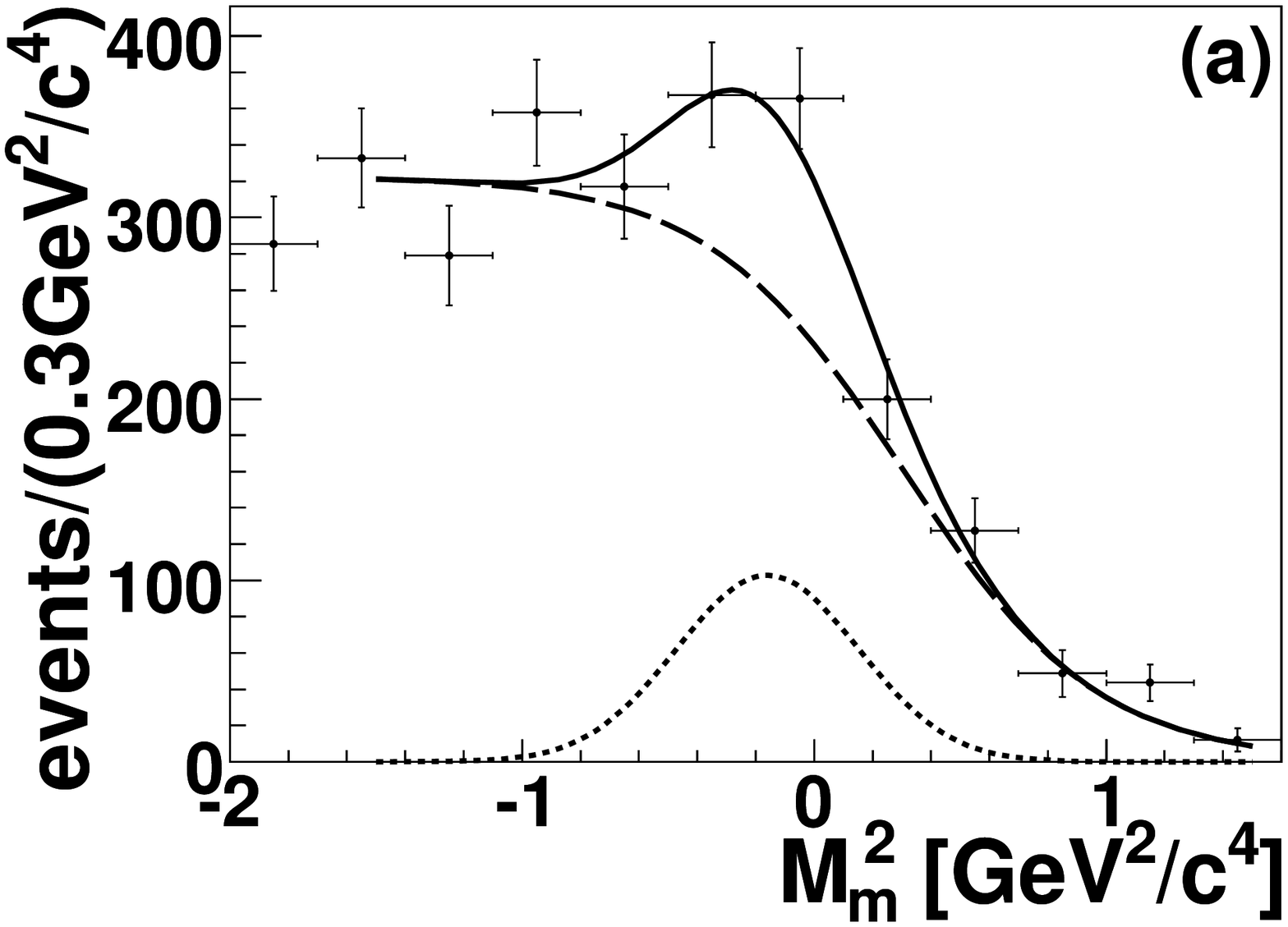}
  \begin{minipage}[b]{0.238\textwidth}
    \includegraphics[width=\textwidth]{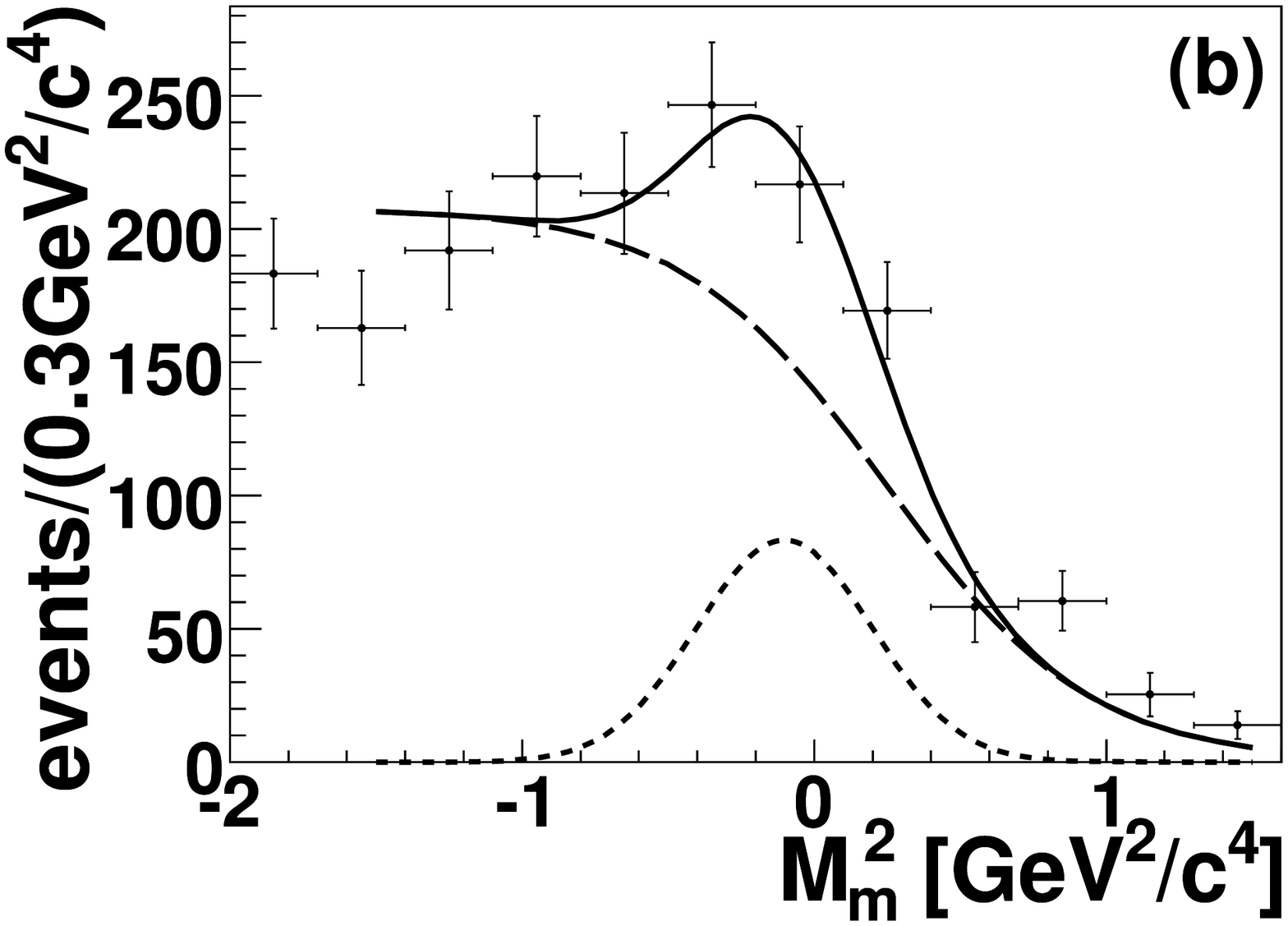}
  \end{minipage}
  \caption{\sl Sideband subtracted \MM\ distributions with fitted functions superimposed: (a) for 	the electron channel and (b) for the muon channel. All \Ds\ reconstruction chains 
    have been summed. Solid lines represent the full distribution, dashed lines are the \Ds\ background component and dotted lines represent the fitted signal component.}
  \label{fig:Signal}
\end{figure}

Fit results are given in Table \ref{tab:fitresults}. Reconstruction efficiencies for the three decay chains are obtained by counting simulated signal events in the range $|\MM|<1.2 \gevccc$. As reported in Ref.~\cite{CLEO_Phi} the reconstruction efficiency of the $\Ds \to \phi \pip$ decay chain depends on the requirement on the $\phi$ mass. The impact of this effect is covered by the systematic uncertainties on $\epsilon_{BR}$. Figure~\ref{fig:Signal} shows the 
sideband subtracted \MM\ distributions summed over the decay channels. 

The bias-corrected signal yields are $N^{\mathrm{signal}}_{\mathrm{electron}}=301^{+68}_{-67}$ and $N^{\mathrm{signal}}_{\mathrm{muon}}=206^{+53}_{-52}$. The bias correction is $+42$ $(-4)$ events for the electron (muon) channel. Extended simulations showed that the source of the bias is a fluctuation of the underlying combinatorial background distribution.

The systematic uncertainties are divided into two categories: additive uncertainties 
(Table \ref{tab:addsystematics}) are related to the number of extracted signal events, while multiplicative uncertainties (Table \ref{tab:multisystematics}) are  related to the calculated branching fraction. 
The uncertainty due to the \Ds\ daughter branching fractions is quoted separately.
\begin{table}[hbt]
  \begin{center}
    \caption[Overview of the systematic uncertainties.]
    {Additive systematic uncertainties in events.}
    \vspace{2pt}
    \label{tab:addsystematics}
    \begin{tabular}{l c c}
      \hline \hline
      source & $\Delta N_{\mathrm{elec.}}$ [Evts]& $\Delta N_{\mathrm{muon}}$ [Evts]\\
      \hline
      \Ds\ bkg parameterization & $19.9$ & $15.9$\\
      Single channel signal yields & $14.5$ & $9.0$\\
      Width of the signal PDF & $3.9$ & $4.3$\\
      Error of the $m(\Ds)$ fits & $3.6$ & $3.4$\\
      \hline
      Total, affecting significance & $25.2$ & $19.1$\\
      \hline
      Bias correction & $2.2$ & $1.8$\\
      \hline
      Total uncertainty & $25.3$ & $19.2$\\
      \hline \hline
    \end{tabular}
  \end{center}
\end{table}

The systematic uncertainty arising from the choice of the \Ds\ background PDF is evaluated using 1000 statistically independent MC experiments. Each experiment corresponds to different values for the two parameters that describe the PDF, \MO\ and \EC, which are distributed according to the error matrix for these parameters. We take the width of a Gaussian fitted to the resulting $N^{\mathrm{signal}}$ distribution as a systematic uncertainty. The impact of shape differences between data and MC have been studied, as well as shape differences due to varying compositions of the \Ds\ background, and both found to be negligible. 
A similar procedure is used to estimate the uncertainty due to using the \Ds\ branching fractions $\epsilon_{BR}$ for the combination of the individual channel signal yields. MC samples of $\epsilon_{BR}$ are produced for each decay channel using the information of Ref.~\cite{PDG}. This leads to differences in the total number of extracted signal events. The width of a Gaussian fitted to the resulting distribution of signal yields is taken as systematic uncertainty.

The width of the Gaussian PDF of \MM\ for signal, as well as the number of fitted \Ds, are varied by $\pm 1\sigma$ to evaluate these systematic uncertainties. This approach also takes into account the variation of the width due to a contribution of \Dss\ to the signal yield. The systematic uncertainty related to the bias correction is given by the statistical uncertainty of the correction.

We evaluate the uncertainty of the signal MC model by calculating the difference of the efficiencies between the alternative signal models and the Goity-Roberts signal MC model. The impact of the finite statistics of the simulated signal sample is deduced from the uncertainty on the efficiency determination. The uncertainty arising from particle identification, as well as from the \KS\ reconstruction, is determined using dedicated high purity control samples for the corresponding particles. Uncertainties arising from track and photon reconstruction, as well as from radiative corrections, are evaluated by varying their reconstruction efficiencies and the energy radiated by photons in the simulation. The uncertainty on the number of \B\ mesons in the data set is determined as described in Ref.~\cite{bcount} and the \Ds\ 
daughter branching fraction uncertainties are taken from Ref.~\cite{PDG}. 
\begin{table}[hbt]
  \begin{center}
    \caption[Overview of the multiplicative systematic uncertainties.]
    {Multiplicative systematic uncertainties in percent.}
    \vspace{2pt}
    \label{tab:multisystematics}
    \begin{tabular}{l c c c}
      \hline \hline
      source & \multicolumn{3}{c}{syst. uncer. electron (muon) ch. $[\%]$}\\

                 & $\phi \pip$ & $\Kstarzb \Kp$ & $\KS \Kp$ \\
      \hline             
      Signal MC model         & $7.6\,(0.2)$ & $3.1\,(6.4)$ & $5.9\,(2.1)$   \\
      N(Signal MC)            & $2.7\,(3.5)$ & $3.3\,(4.2)$ & $2.5\,(3.3)$   \\
      Particle ID           & $0.6\,(1.6)$ & $1.2\,(4.9)$ & $3.6\,(7.9)$   \\
      \KS\ eff.            & - (-) & - (-) & $2.0\,(3.1)$           \\
      Tracking eff.             & $0.4\,(0.1)$ & $0.5\,(0.2)$ & $1.8\,(2.4)$   \\
      Photon eff.          & $0.6\,(0.9)$ & $0.4\,(0.9)$ & $0.5\,(0.7)$   \\
      Radiative corr.       & $2.0\,(2.1)$ & $2.2\,(2.5)$ & $1.9\,(1.9)$   \\
      \hline
      Total $\Delta \epsilon_{\mathrm{reco}}$  & $8.4\,(4.5)$ & $5.2\,(9.5)$ & $8.1\,(9.9)$   \\
      \hline
      $B$ counting          & $1.1\,(1.1)$ & $1.1\,(1.1)$ & $1.1\,(1.1)$   \\
      \BR(\Ds)              & $15.1\,(15.1)$ & $15.4\,(15.4)$ & $6.0\,(6.0)$ \\
      \hline \hline
    \end{tabular}
  \end{center}
\end{table}

A second fit, imposing an $N^{\mathrm{signal}} = 0$ hypothesis, is used to estimate the significance of the measurement. 
Since the mean of the Gaussian is a free parameter in the signal PDF, the difference in the number of free parameters ($\Delta NDF$) of the fits is larger than one. As shown in Ref.~\cite{Statistics}, the resulting probability distribution cannot be approximated by a chisquare distribution with an integer number of degrees of freedom. Thus, only a significance range, representing the significances for $\Delta NDF=2$ and $\Delta NDF=1$, is calculated.
Including statistical and systematic uncertainties, the ranges are $[3.3\, -\, 3.7]\sigma$ and $[3.5\, - \, 3.9]\sigma$ for the electron and muon channel, respectively. Combining both lepton channels results in a significance larger than $5.0\sigma$. 

The branching fractions for the individual lepton channels are 
$\BR(\BtoDsKenuX)=[5.81^{+1.30}_{-1.30}(\mathrm{stat.})\pm0.54(\mathrm{syst.})\pm0.49(\BR(\DsO))]\times10^{-4}$
and 
$\BR(\BtoDsKmunuX)=[6.68^{+1.72}_{-1.69}(\mathrm{stat.})\pm0.69(\mathrm{syst.})\pm0.56(\BR(\DsO))]\times10^{-4}$,
where the last uncertainty reflects 
the limited knowledge of the \DsO\ branching fractions. The measurements are combined, taking into account the correlations between their systematic uncertainties, yielding $\BR(\BtoDsKlnuX) = [6.13^{+1.04}_{-1.03}(\mathrm{stat.})\pm0.43(\mathrm{syst.}) \pm 0.51(\BR(\DsO))]\times10^{-4}$.

In summary, using a data sample of about $376.9$ million \BB\ pairs, we find evidence for the 
decay \BtoDsKlnuX. The signal has a significance larger than $5.0 \sigma$, after taking systematic effects into account. The measured branching fraction, $\BR(\BtoDsKlnuX) = [6.13^{+1.04}_{-1.03}(\mathrm{stat.})\pm0.43(\mathrm{syst.}) \pm 0.51(\BR(\DsO))]\times10^{-4}$, where the last uncertainty reflects 
the limited knowledge of the \DsO\ branching fractions, is consistent with the previous upper limit reported by the ARGUS collaboration and with theoretical expectations. 

We are grateful for the excellent luminosity and machine conditions
provided by our \pep2\ colleagues 
and for the substantial dedicated effort from
the computing organizations that support \babar.
The collaborating institutions wish to thank 
SLAC for its support and kind hospitality. 
This work is supported by
DOE
and NSF (USA),
NSERC (Canada),
CEA and
CNRS-IN2P3
(France),
BMBF and DFG
(Germany),
INFN (Italy),
FOM (The Netherlands),
NFR (Norway),
MES (Russia),
MICIIN (Spain),
STFC (United Kingdom). 
Individuals have received support from the
Marie Curie EIF (European Union),
the A.~P.~Sloan Foundation (USA)
and the Binational Science Foundation (USA-Israel).


\begin{thebibliography}{99}
\bibitem{conjugate}
  Throughout this paper, whenever a mode is given, the charge
  conjugate is implied.
\bibitem{CMframe}
  Throughout this paper, all kinematic variables are defined in the 
  center of mass frame.
\bibitem{Legendre}
  $L_{2} = \sum_i|\vec{p}_{i}| cos^{2}(\theta_{i})$, where the sum is over all tracks in the event not used
to form the signal \B\ candidate, and $\vec{p}_{i}$ and $\theta_i$ are the momenta and angles with respect to the thrust axis of the signal \B\ candidate, respectively.
\end{thebibliography}
\end{document}